\documentclass[journal=jctcce,manuscript=article]{achemso}

\usepackage{chemformula} 
\usepackage[T1]{fontenc} 
\usepackage{amsmath,amssymb,upgreek,multirow,tabularx,booktabs}
\usepackage{graphicx}

\author{Xinqiang Ding}
\email{Xinqiang.Ding@tufts.edu}
\affiliation{Department of Chemistry, Tufts University, 62 Talbot Avenue, Medford, MA 02155}

\title{Bayesian Multistate Bennett Acceptance Ratio Methods}

\abbreviations{BAR, MBAR, BayesMBAR}

\begin{document}

\begin{tocentry}
\includegraphics{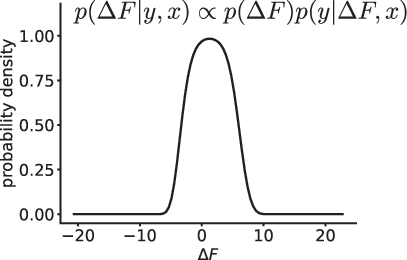}  
\end{tocentry}

\begin{abstract}
The multistate Bennett acceptance ratio (MBAR) method is a prevalent approach for computing free energies of thermodynamic states.
In this work, we introduce BayesMBAR, a Bayesian generalization of the MBAR method.
By integrating configurations sampled from thermodynamic states with a prior distribution, BayesMBAR computes a posterior distribution of free energies.
Using the posterior distribution, we derive free energy estimations and compute their associated uncertainties.
Notably, when a uniform prior distribution is used, BayesMBAR recovers the MBAR's result but provides more accurate uncertainty estimates.
Additionally, when prior knowledge about free energies is available, BayesMBAR can incorporate this information into the estimation procedure by using non-uniform prior distributions.
As an example, we show that, by incorporating the prior knowledge about the smoothness of free energy surfaces, BayesMBAR provides more accurate estimates than the MBAR method.
Given MBAR's widespread use in free energy calculations, we anticipate BayesMBAR to be an essential tool in various applications of free energy calculations.
\end{abstract}

\section{Introduction}
Computing free energies of thermodynamic states is a central problem in computational chemistry and physics.
It has wide-ranging applications including computing protein-ligand binding affinities \cite{choderaAlchemicalFreeEnergy2011}, predicting molecular solubilities\cite{mobleySmallMoleculeHydration2009}, and estimating phase equilibria\cite{athanassioszpanagiotopoulosMonteCarloMethods2000,dybeckCapturingEntropicContributions2017,schieberUsingReweightingFree2018}, among other tasks.
For states whose free energies are not analytically tractable, their free energies are often estimated using numerical methods\cite{chipotFreeEnergyCalculations2007}.
These methods typically involve sampling configurations from states of interest and subsequently computing their free energies based on sampled configurations.
In this work we focus on the second step of estimating free energies, assuming
that equilibrium configurations have been sampled using Monte Carlo sampling or molecular dynamics.

The multistate Bennett acceptance ratio (MBAR) method\cite{shirtsStatisticallyOptimalAnalysis2008} is a common technique for estimating free energies given sampled configurations.
Equivalent formulations of the MBAR method were also developed in different contexts\cite{tanTheoryBinlessMultistate2012,kongTheoryStatisticalModels2003,geyerEstimatingNormalizingConstants1994}.
For the purpose of this study, we refer to this method and its equivalent formulations as MBAR.
The MBAR method not only offers an estimate of free energies but also provides the statistical uncertainty associated with the estimate.
In situations where a large number of configurations are available, the MBAR estimator is unbiased and has the smallest variance among estimators reliant on sampled configurations. \cite{kongTheoryStatisticalModels2003,shirtsStatisticallyOptimalAnalysis2008}
However, properties of the MBAR estimator and their associated uncertainty estimate remain largely unexplored when the number of configurations is small. 
Furthermore, in such scenarios, it becomes desirable to incorporate prior knowledge into the estimation procedure.

A systematic approach of integrating prior knowledge into an estimation procedure is Bayesian inference\cite{bergerStatisticalDecisionTheory2013}. 
Bayesian inference treats unknown quantities (free energies in this case) as random variables and incorporates prior knowledge into the estimation procedure by employing prior distributions and the Bayes' theorem. 
In terms of free energy estimation, prior knowledge could come from previous simulations, experiments, or physical knowledge of a system.
A common instance of physical prior knowledge on free energies is that free energy surfaces along a collective coordinate are usually smooth.
Combining prior knowledge with observed data (configurations sampled from thermodynamic states), Bayesian inference computes the posterior distribution of the unknown quantities.
The posterior distribution provides both estimates of the unknown quantities and the uncertainty of the estimates.

Estimating free energies using Bayesian inference has been investigated in multiple studies.
For instance, Stecher et al.\cite{stecherFreeEnergySurface2014} used a Gaussian process as the prior distribution over smooth free energy surfaces. 
The resulting posterior distribution, given configurations from umbrella sampling, was utilized to estimate free energy surfaces and associated uncertainty. 
Shirts et al.\cite{shirtsStatisticallyOptimalContinuous2020} parameterized 
free energy surfaces using splines and constructed prior distributions using a Gaussian prior on spline coefficients.
Unlike these studies that primarily focused on estimating free energy surfaces from biased simulations, the works of Habeck \cite{habeckBayesianReconstructionDensity2007,habeckBayesianEstimationFree2012},Ferguson \cite{fergusonBayesWHAMBayesianApproach2017}, and Maragakis et al.\cite{maragakisBayesianEstimatesFree2008} were aimed at estimating densities of states and free energy differences using Bayesian inference.
Methods developed in these studies are direct Bayesian generalizations of the weighted histogram analysis method (WHAM) and the Bennett acceptance ratio (BAR) method \cite{bennettEfficientEstimationFree1976}.


This work focuses on improving the accuracy of estimating free energies of discrete thermodynamic states when the number of sampled configurations is small.
For this purpose, we developed a Bayesian generalization of the MBAR method, which we term BayesMBAR.
With several benchmark examples, we show that, when the number of configurations is small, BayesMBAR provides not only superior uncertainty estimates compared to MBAR but also more accurate estimates of free energies by incorporating prior knowledge into the estimation procedure.

\section{Methods}
The MBAR method is commonly understood as a set of self-consistent equations, which is not amenable to the development of its Bayesian generalization.
To develop a Bayesian generalization of MBAR, we first emphasize the probabilistic nature of the MBAR method.
Although there are multiple statistical models from which the MBAR method can be derived, we build upon the reverse logistic regression model\cite{geyerEstimatingNormalizingConstants1994}, which treats free energies as parameters and provides a likelihood for inference.
To convert the reverse logistic regression model into a Bayesian model, we treat free energies as random variables and place a prior distribution on them.
Then the posterior distribution of free energies is computed using the Bayes' theorem.
Samples from the posterior distribution are efficiently generated using Hamiltonian Monte Carlo (HMC) methods \cite{nealMCMCUsingHamiltonian2011,hoffmanNoUturnSamplerAdaptively2014}.
These samples are used to estimate free energies and quantify the
uncertainty of the estimate.
Hyperparameters of the prior distribution are automatically optimized by maximizing the marginal likelihood of data (Bayesian evidence).
We present the details of BayesMBAR in the following sections.

\subsection*{The reverse logistic regression model of MBAR}
Computing free energies of thermodynamic states is closely related to computing normalizing constants of Bayesian models.
Multiple methods have been developed in statistics for estimating normalizing constants \cite{geyerEstimatingNormalizingConstants1994,johnskillingNestedSamplingGeneral2006,mengWarpBridgeSampling2002,mengSIMULATINGRATIOSNORMALIZING1996} and these methods are directly applicable for estimating free energies.
Here we focus on the reverse logistic regression method proposed by Geyer\cite{geyerEstimatingNormalizingConstants1994} and show that the solution of this method is equivalent to the MBAR method.


Let us assume that we aim to calculate free energies of $m$ thermodynamic states (up to an additive constant) by sampling their configurations.
Let $u_i(x), i = 1, ..., m$ be the reduced potential energy functions\cite{shirtsStatisticallyOptimalAnalysis2008} of the $m$ states.
The free energy of the $i$th state is defined as  
\begin{equation}
  F_i = - \log \int_{\Upgamma}{e^{-u_i(x)}} \mathrm{d}x,
\end{equation}
where $\Upgamma$ is the configuration space.
For the $i$th state, $n_i$ uncorrelated configurations, $\{x_{ik}, k=1, ..., n_i\}$, are sampled from its Boltzmann distribution $p_i(x; F_i) = \exp(-[u_i(x) - F_i])$.
Here $p_i(x; F_i)$ means that it is a distribution of the random variable $x$ with parameter $F_i$. 
We will henceforth use such notation of separating parameters from random variables with a semicolon in probability distributions.
To estimate free energies, Geyer\cite{geyerEstimatingNormalizingConstants1994} proposed the following retrospective formulation.
This formulation treats indices of states in an unconventional manner.
Let us use $y_{ik}$ to denote the index of the state from which configuration $x_{ik}$ is sampled.\cite{geyerEstimatingNormalizingConstants1994}
Apparently, $y_{ik} = i$ for all $i$ and $k$.
Although indices of states for sampled configurations are determined in the sampling setup, they are treated as a multinomial distributed random variable with parameters $\pi = (\pi_1, ..., \pi_m)$.
Because $n_i$ configurations are sampled from state $i$,  the maximum likelihood estimate of $\pi_i$ is $\hat{\pi}_i = n_i/n$, where $n = \sum_{i=1}^{m}n_i$.
The concatenation of state indices and configurations, $(y, x)$, is viewed as samples from the joint distribution of $p(y, x)$, which is defined as
\begin{align}
    p(y = i, x; F_i, \log \pi_i) & = p(y=i) \cdot p(x|y=i; F_i)\\
    & = e^{-[u_i(x) - F_i - \log \pi_i]},
\end{align}
for $i \in \{1, ..., m\}$.
Following a retrospective argument, the reverse logistic regression method  estimates the free energies by asking the following question.
Given that a configuration $x$ is observed, what is the probability that it is sampled from state $y=i$ rather than other states?
Using Bayes' theorem, we can compute this retrospective conditional probability as
\begin{align}
  \label{eq:conditional}
    p(y=i|x; F, \log \pi) &= \frac{p(y=i, x; F_i, \log \pi_i)}{\sum_{j=1}^{m}p(y=j, x; F_j, \log \pi_j)} = \frac{e^{-[u_i(x) - F_i - \log \pi_i]}}{\sum_{j=1}^{m} e^{-[u_j(x) - F_j - \log \pi_j]}},
\end{align}
where $F = (F_1, ..., F_m)$ and $\log \pi = (\log \pi_1, ..., \log \pi_m)$.
The free energies are estimated by maximizing the product of the retrospective conditional probabilities of all configurations, which is equivalent to maximizing the log-likelihood
\begin{align}
	\label{eq:log-likelihood}
    \ell(F, \log \pi) &= \sum_{i=1}^{m}\sum_{k=1}^{n_i} \log p(y_{ik} = i| x_{ik}; F, \log \pi) \nonumber \\
    &= \sum_{i=1}^{m}\sum_{k=1}^{n_i} \Big[ -[u_i(x_{ik}) - F_i - \log \pi_i] - \log \sum_{j=1}^{m} e^{-[u_j(x_{ik}) - F_j - \log \pi_j]} \Big].
\end{align}
The log-likelihood function $\ell(F, \log \pi)$ in Eq. \ref{eq:log-likelihood} depends on $F$ and $\log \pi$ only through their sum $\phi = F + \log \pi$, so $F$ and $\log \pi$ are not separately estimable from maximizing the log-likelihood.
The solution is to substitute $\log \pi_i$ with the empirical estimate $\log \hat{\pi}_i = \log (n_i/n)$. 
Then setting the derivative of $\partial \ell/\partial F$ to zero, we obtain 
\begin{align}
	\label{eq:mbar}
    \hat{F}_r = -\log \sum_{i=1}^{m} \sum_{k=1}^{n_i} \frac{e^{-u_r(x_{ik})}}{\sum_{j=1}^{m} n_j e^{-[u_j(x_{ik}) - \hat{F}_j]}}.
\end{align}
for $r = 1, ..., m$.
$\hat{F} = (\hat{F}_1, ..., \hat{F}_m)$ is the solution that maximizes $\ell(F, \log \hat{\pi})$.
Eq. \ref{eq:mbar} is identical to the MBAR equation and reduces to the BAR equation \cite{bennettEfficientEstimationFree1976} when $m = 2$.
The technique described above is termed ``reverse logistic regression'' based on two primary insights. 
First, the log-likelihood in equation \ref{eq:log-likelihood} bears resemblance to that found in multi-class logistic regression. 
Second, the primary goal of this method is to estimate $F$, the intercept term. 
This differs from traditional logistic regression, where the aim is to determine regression coefficients and predict the response variable $y$.

The uncertainty of the estimate $\hat{F}$ is computed using asymptotic analysis of the log-likelihood function $\ell(F, \log \pi)$ in Eq. \ref{eq:log-likelihood}.
Because the log-likelihood function $\ell(F, \log \pi)$ depends on $F$ and $\log \pi$ only through their sum $\phi =  F + \log \pi$, the observed Fisher information matrix computed using the log-likelihood function can only be used to compute the asymptotic covariance of the sum $\phi$.
The observed Fisher information matrix at $\hat{\phi} = \hat{F} + \log \hat{\pi} $ is
\begin{align}
  \mathcal{J}_\phi =  \sum_{i=1}^{m}\sum_{k=1}^{n_i} (\mathrm{diag}(p_{ik}) - p_{ik} p_{ik}^\top),
\end{align}
where $p_{ik}$ is a column vector of $(p(y_{ik} = 1|x_{ik}; \hat{\phi}), ..., p(y_{ik} = m|x_{ik}; \hat{\phi}) )$, and $\mathrm{diag}(p_{ik})$ is a diagonal matrix with $p_{ik}$ as its diagonal elements.
The asymptotic covariance matrix of $\hat{\phi}$ is the Moore-Penrose pseudo-inverse of the observed Fisher information matrix, i.e., $\mathrm{cov}(\hat{\phi}) = \mathcal{J}_\phi^{-}$.
To compute the asymptotic covariance matrix of $\hat{F}$, we assume that $\hat{\phi}$ and $\log \hat{\pi}$ are asymptotically independent.
Then the asymptotic covariance matrix of $\hat{F}$ can be computed as 
\begin{align}
  \label{eq:asymptotic}
  \mathrm{cov}(\hat{F}) &= \mathrm{cov}(\hat{\phi}) - \mathrm{cov}(\log \hat{\pi}) \nonumber \\
  &= \mathcal{J}_\phi^{-} - \mathrm{diag}(1/(n\hat{\pi})) + \mathbf{1} \mathbf{1}^\top / n,
\end{align}
where $\mathbf{1}$ is a column vector of $m$ ones.
The asymptotic covariance matrix in Eq. \ref{eq:asymptotic} is the same as that derived in Ref. \cite{kongTheoryStatisticalModels2003} and is commonly used in the MBAR method \cite{shirtsStatisticallyOptimalAnalysis2008}.
With the asymptotic covariance matrix of $\hat{F}$, we can compute the asymptotic variance of their differences using the identity 
$\mathrm{var}(\hat{F}_r - \hat{F}_s) = \mathrm{cov}(\hat{F}_r, \hat{F}_r) + \mathrm{cov}(\hat{F}_s, \hat{F}_s) - 2 \cdot \mathrm{cov}(\hat{F}_r, \hat{F}_s)$, where $\mathrm{cov}(\hat{F}_r, \hat{F}_s)$ is the $(r, s)$th element of the matrix $\mathrm{cov}(\hat{F})$.


\subsection*{Bayesian MBAR}
As shown above, the reverse logistic regression model formulates MBAR as a statistical model.
It provides a likelihood function (Eq. \ref{eq:log-likelihood}) for computing the MBAR estimate $\hat{F}$ and the associated asymptotic covariance.
Based on this formulation, we developed BayesMBAR by turning the reverse logistic regression into a Bayesian model.
In BayesMBAR, we treat $F$ as a random variable instead of a parameter and place a prior distribution on $F$.
The posterior distribution of $F$ is then used to estimate $F$.
Let us represent the prior distribution of $F$ as $p(F;\theta)$, where $\theta$ is the parameters of the prior distribution and is often called hyperparameters.
Borrowing from the reverse logistic regression, we use the retrospective conditional probability in  Eq. \ref{eq:conditional} as the likelihood function, i.e., $p(y|x, F) = p(y|x; F, \log \pi)$.
We note that $F$ is treated as a random variable in $p(y|x, F)$ whereas it is a parameter in $p(y|x; F, \log \pi)$.
The $\log \pi$ term in the likelihood function is substituted with the maximum likelihood estimate $\log \hat{\pi}$.
With these definitions, the posterior distribution of $F$ given sampled configurations and state index is
\begin{align}
  \label{eq:posterior}
    p(F|Y,X) &= \frac{p(Y|F, X) p(F;\theta)}{\int p(Y|F, X) p(F;\theta) dF} \nonumber \\
    & \propto p(F; \theta) \prod_{i=1}^{m}\prod_{k=1}^{n_i} p(y_{ik}|x_{ik}; F),
\end{align}
where $Y = \{ y_{ik}: i = 1, ..., m; k = 1, ..., n_i \}$ and $X = \{ x_{ik}: i = 1, ..., m; k = 1, ..., n_i \}$.
Using the posterior distribution in Eq. \ref{eq:posterior}, we can compute various quantities of interest such as the posterior mode and the posterior mean, both of which can serve as point estimates of $\hat{F}$.
In addition, we can use the posterior covariance matrix as an estimate of the uncertainty for $\hat{F}$.
However, to carry out these calculations, we need to address the following questions that commonly arise in Bayesian inference.

\subsection*{Choosing the prior distribution}
To fully specify the BayesMBAR model, we need to choose a prior distribution for $F$.
We could use information about $F$ from previous simulations or experiments to construct the prior distribution if such information is available.
For example, the prior distribution of protein-ligand binding free energies could be constructed using free energies computed with fast but less accurate methods such as docking.
The information could also come from the binding free energies of similar protein-ligand systems.
Turning such information into a prior distribution will depend on domain experts' experience and likely vary from case to case.
In this work, we focus on scenarios where such information is not available.
In this scenario, we propose to use two types of distributions as the prior: the uniform distribution and the Gaussian distribution.

\textbf{Using uniform distributions as the prior.}
As the MBAR method has proven to be a highly effective method for estimating free energies in many applications, a conservative strategy for choosing the prior distribution is to minimize the deviation of BayesMBAR from MBAR.
Such a strategy leads to using the uniform distribution as the prior distribution, because it makes the maximum a posteriori probability (MAP) estimate of BayesMBAR the same as the MBAR estimate.
Specifically, if we set the prior distribution of $F$ to be the uniform distribution, i.e., $p(F; \theta) \propto \mathrm{constant}$, the posterior distribution of $F$ in  Eq. \ref{eq:posterior} becomes the same as the likelihood function.
Therefore, maximizing the posterior distribution of $F$ is equivalent to maximizing the log-likelihood function in Eq. \ref{eq:log-likelihood}.

While recovering the MBAR estimate with its MAP estimate, 
BayesMBAR with a uniform prior distribution provides two advantages.
First, in addition to the MAP estimate, BayesMBAR also offers the posterior mean as an alternative point estimate of $F$.
Second, BayesMBAR produces a posterior distribution of $F$, which can be used to estimate the uncertainty of the estimate.
As shown in the Result sections, the uncertainty estimate from BayesMBAR is more accurate than that from MBAR when the number of configurations is small.

\textbf{Using Gaussian distributions as the prior.}
In many applications, we are interested in computing free energies along collective coordinates such as distances, angles or alchemical parameters.
In such cases, we often have the prior knowledge that the free energy surface is a smooth function $F(\lambda)$ of the collective coordinate $\lambda$.
A widely used approach to encode such knowledge into Bayesian inference is to use a Gaussian process\cite{rasmussenGaussianProcessesMachine2005} as the prior distribution.
A Gaussian process is a collection of random variables, any finite number of which have a joint Gaussian distribution.
A Gaussian process is fully specified by its mean function $\mu(\lambda)$ and covariance function $k(\lambda, \lambda^\prime)$.
The value of the covariance function $k(\lambda, \lambda^\prime)$ is the covariance between $F(\lambda)$ and $F(\lambda^\prime)$.
The covariance function is often designed to encode the smoothness of the function.
Specifically, the covariance $k(\lambda, \lambda^\prime)$ between $F(\lambda)$ and $F(\lambda^\prime)$ increases as $\lambda$ and $\lambda^\prime$ become closer.
When the mean function is smooth and a covariance function such as the squared exponential covariance function is used, the Gaussian process is a probability distribution of smooth functions.

In BayesMBAR we focus on estimating free energies at discrete values of the collective coordinate, $(\lambda_1, ..., \lambda_m)$, instead of the whole free energy surface.
Projecting the Gaussian process over free energy surfaces onto discrete values of $\lambda$, we obtain as the prior distribution of $F = (F(\lambda_1), ..., F(\lambda_m))$ a multivariate Gaussian distribution with the mean vector $\mu = (\mu(\lambda_1), ..., \mu(\lambda_m))$ and the covariance matrix $\Sigma$.
The $(i,j)$th element of $\Sigma$ is computed as $\Sigma_{ij} = k(\lambda_i, \lambda_j)$ and represents the covariance between $F(\lambda_i)$ and $F(\lambda_j)$.
As in many applications of Gaussian processes, we set the mean function to be a constant function, i.e., $\mu = (c, ..., c)$, where $c$ is a hyperparameter to be optimized.
The choice of the covariance function is a key ingredient of constructing the prior distribution, as it encodes our assumption about the free energy surface's smoothness.\cite{rasmussenGaussianProcessesMachine2005}
Several well-studied covariance functions are suitable for use in BayesMBAR.
In this study we use the squared exponential covariance function as an example, noting that other types of covariance function can be used as well.
The squared exponential covariance function is defined as
\begin{align}
  k_{\mathrm{SE}}(\lambda, \lambda^\prime) = \sigma^2 \cdot \exp \Big( -\frac{r^2}{2l^2} \Big),
\end{align}
where $r = |\lambda - \lambda^\prime|$ and the variance scale $\sigma$ and the length scale $l$ are hyperparameters to be optimized.
Every function $F(\lambda)$ from such Gaussian processes has infinitely many derivatives and is very smooth.
The hyperparameters $\sigma$ and $l$ control the variance and the length scale of the function $F(\lambda)$, respectively.
The collective coordinate $\lambda$ is not restricted to scalars and can be a vector.
When $\lambda$ is a vector, the length scale $l$ could be either a scalar or a vector of the same dimension as $\lambda$, the latter of which means that the length scale can be different for different dimensions in $\lambda$.
When a vector length scale is used, the squared exponential covariance function is defined as
\begin{align}
  k_{\mathrm{SE}}(\lambda, \lambda^\prime) = \sigma^2 \cdot \exp \Big( -\frac{1}{2} (\lambda - \lambda^\prime)^\top L^{-1} (\lambda - \lambda^\prime) \Big),
\end{align}
where $L$ is a diagonal matrix with values of $l$ as its diagonal elements.
To allow the variances to be different for different entries in $F = (F(\lambda_1), ..., F(\lambda_m))$, we add a constant $\tilde{\sigma}_i^2$ to the variance of $F(\lambda_i)$, i.e., $\Sigma_{ii} = k(\lambda_i, \lambda_i) = k_{\mathrm{SE}}(\lambda_i, \lambda_i) + \tilde{\sigma}_i^2$.
With the mean function and the covariance function defined, the prior distribution of $F$ is fully specified as a multivariate Gaussian distribution of 
\begin{align}
  \label{eq:prior}
  p(F; \theta) = \frac{1}{(2\pi)^{m/2} |\Sigma_\theta|^{1/2}} \exp(-\frac{1}{2} (F - \mu_\theta)^T \Sigma_\theta^{-1} (F - \mu_\theta)),
\end{align}
where $\mu_\theta$ and $\Sigma_\theta$ are the mean vector and the covariance matrix, respectively. They depend on the hyperparameters $\theta = (c, \sigma, \tilde{\sigma}, l)$, where $\tilde{\sigma} = (\tilde{\sigma}_1, ..., \tilde{\sigma}_m)$.
The hyperparameters $\theta$ are optimized by maximizing the Bayesian evidence, as described in following sections.

\subsection*{Computing posterior statistics.}
With the prior distribution of $F$ defined as above, the posterior distribution defined in Eq. \ref{eq:posterior} contains rich information about $F$.
Specifically, the MAP or the posterior mean can be used as point estimates of $F$ and the posterior covariance matrix can be used to compute the uncertainty of the estimate.

\textbf{Computing the MAP estimate.}
The MAP estimate of $F$ is the value that maximizes the posterior distribution density, i.e., $\hat{F} = \underset{F}{\arg\max} \log p(F|Y,X)$.
When the prior distribution is chosen to be either uniform distributions or the Gaussian distributions, $\log p(F|Y,X)$ is a concave function of $F$.
This means that the MAP estimate is the unique global maximum of the posterior distribution density and can be efficiently computed using standard optimization algorithms.
In BayesMBAR, we implemented the L-BFGS-B algorithm\cite{liuLimitedMemoryBFGS1989} and the Newton's method to compute the MAP estimate.

\textbf{Computing the mean and the covariance matrix of the posterior distribution.}
Computing the posterior mean and the covariance matrix is more challenging than computing the MAP estimate. 
It involves computing an integral with respect to the posterior distribution density.
When there are only two states, we compute the posterior mean and the covariance matrix by numerically integration.
When there are more than two states, numerical integration is not feasible.
In this case, we estimate the posterior mean and covariance matrix by sampling from the posterior distribution using the No-U-Turn Sampler (NUTS).\cite{hoffmanNoUturnSamplerAdaptively2014}
The NUTS sampler is a variant of Hamiltonian Monte Carlo (HMC) methods \cite{nealMCMCUsingHamiltonian2011} and has the advantage of automatically tuning the step size and the number of steps.
The NUTS sampler has been shown to be highly efficient in sampling from high-dimensional distributions for Bayesian inference problems.\cite{carpenterStanProbabilisticProgramming2017,pmlr-v118-xu20a}
In BayesMBAR, an extra factor that further improves the efficiency of the NUTS sampler is that the posterior distribution density is a concave function, which means that the sampler does not need to cross low density (high energy) regions during sampling.
In BayesMBAR, we use the NUTS sampler as implemented in the Python package, \texttt{BlackJAX}\cite{laoBlackjaxSamplingLibrary2020}.

\subsection*{Optimizing hyperparameters}
When Gaussian distributions with a specific covariance function are used as the prior distribution of $F$ (Eq. \ref{eq:prior}), we need to make decisions about the values of hyperparameters.
Such decisions are referred to as model selection problems in Bayesian inference and several principles have been proposed and used in practice.
In BayesMBAR, we use the Bayesian model selection principle, which is to choose the model that maximizes the marginal likelihood of the data.
The marginal likelihood of the data is also called the Bayesian evidence and is defined as
\begin{align}
  \label{eq:marginal}
  p(Y|X; \theta) = \int p(Y|F, X) p(F; \theta) dF.
\end{align}
Because the Bayesian evidence is a multidimensional integral, computing it with numerical integration is not feasible.
In BayesMBAR, we use ideas from variational inference\cite{jordanIntroductionVariationalMethods1999} and Monte Carlo integration\cite{kongTheoryStatisticalModels2003} to approximate it and optimize the hyperparameters.

We introduce a variational distribution $q(F)$ and use the evidence lower bound (ELBO) of the marginal likelihood as the objective function for optimizing the hyperparameters.
Specifically, the ELBO is defined as
\begin{align}
  \label{eq:elbo}
  \mathcal{L}(q, \theta) &= \int q(F) \log \frac{p(Y|F, X) p(F; \theta)}{q(F)} dF \nonumber \\
  &= \mathop{\mathbb{E}}_{F\sim q} \big[ \log p(Y|F, X) + \log p(F; \theta) - \log q(F) \big].
\end{align}
It is straightforward to show that 
$\mathcal{L}(q, \theta) = \log p(Y|X; \theta) - D_{KL}(q||p(F|Y,X; \theta)) \leq \log p(Y|X; \theta)$, where $D_{KL}(q||p(F|Y,X; \theta))$ is the Kullback-Leibler divergence between $q(F)$ and $p(F|Y,X; \theta)$.
Therefore the ELBO is a lower bound of the log marginal likelihood of data and the gap between them is the Kullback-Leibler divergence between $q(F)$ and $p(F|Y,X; \theta)$.
This suggests that, to make the ELBO a good approximation of the log marginal likelihood, we should choose $q(F)$ that is close to $p(F|Y,X; \theta)$.

Although we could in principle use $p(F|Y,X; \theta)$ as the variational distribution $q(F)$ (then the ELBO would be equal to the log marginal likelihood), it is not practical because computing the gradient of the ELBO with respect to the hyperparameters would require sampling from $p(F|Y,X; \theta)$ at every iteration of the optimization and is computationally too expensive.
Instead we choose $q(F)$ to be a Gaussian distribution to approximate the posterior distribution based on the following observations.
The posterior distribution density $p(F|Y,X; \theta)$ is equal to the product of the likelihood function $p(Y|F,X)$ and the prior distribution $p(F; \theta)$ up to a normalization constant.
The likelihood term $p(Y|F,X)$ is a log-concave function of $F$ and does not depend on $\theta$, so we can approximate it using a fixed Gaussian distribution $\mathcal{N}(\mu_0, \Sigma_0)$, where the mean $\mu_0$ and the covariance matrix $\Sigma_0$ are computed by sampling $F$ from $p(Y|F,X)$ once.
Because the prior distribution $p(F; \theta)$ is also a Gaussian distribution, $\mathcal{N}(\mu_\theta, \Sigma_\theta)$, multiplying the fixed Gaussian distribution $\mathcal{N}(\mu_0, \Sigma_0)$ with the prior yields another Gaussian distribution $\mathcal{N}(\mu_q, \Sigma_q)$, where $\mu_q$ and $\Sigma_q$ can be analytically computed as
\begin{align}
  \label{eq:q}
  \mu_q &= \Sigma_q \big( \Sigma_0^{-1} \mu_0 + \Sigma_\theta^{-1} \mu_\theta \big) \\
  \Sigma_q &= \big( \Sigma_0^{-1} + \Sigma_\theta^{-1} \big)^{-1}.
\end{align}
Therefore we choose the proposal distribution $q(F)$ to be the Gaussian distribution  $\mathcal{N}(\mu_q, \Sigma_q)$, where $\mu_q$ and $\Sigma_q$ are computed as above and depend on $\theta$ analytically.
We compute the ELBO and its gradient with respect to $\theta$ using the reparameterization trick\cite{kingmaAutoencodingVariationalBayes2022}.
Specifically, we reparameterize the proposal distribution $q(F)$ using $F = \mu_q + \Sigma_q^{1/2} \epsilon$, where $\epsilon$ is a random variable with the standard Gaussian distribution.
The ELBO can then be written as
\begin{align}
  \mathcal{L}(\theta) = \mathop{\mathbb{E}}_{\epsilon \sim \mathcal{N}(\mathbf{0}, \mathbf{I})} 
  \big[ \log p(Y|\mu_q + \Sigma_q^{1/2} \epsilon, X) \big]
  - \mathrm{D}_{\mathrm{KL}}(\mathcal{N}(\mu_q, \Sigma_q) || \mathcal{N}(\mu_\theta, \Sigma_\theta)).
\end{align}
The first term on the right hand side can be estimated by sampling $\epsilon$ from the standard Gaussian distribution and evaluating the log-likelihood $p(Y|\mu_q + \Sigma_q^{1/2} \epsilon, X)$.
The second term can be computed analytically.
The gradient of the ELBO with respect to $\theta$ are computed using automatic differentiation\cite{bradburyJAXComposableTransformations2018}.


\section{Results}
\subsection*{Computing the free energy difference between two harmonic oscillators.}
We first tested the performance of BayesMBAR by computing the free energy difference between two harmonic oscillators.
In this case, because there are only two states, BayesMBAR reduces to a Bayesian generalization of the  BAR method and we use BayesBAR to refer to it.
The two harmonic oscillators are defined by the potential energy functions of $u_1(x) = \frac{1}{2} k_1 x^2$ and $u_2(x) = \frac{1}{2} k_2 (x - 1)^2$, where $k_1$ and $k_2$ are the force constants and $u_1$ and $u_2$ are in the unit of $k_BT$.
The objective is to compute the free energy difference between them, i.e., $\Delta F = F_2 - F_1$ and $F_i = -\log \int e^{-u_i(x)} \mathrm{d}x$ for $i = 1$ and $2$.

We first draw $n_1$ and $n_2$ samples from the Boltzmann distribution of $u_1$ and $u_2$, respectively.
Then we use BayesBAR with the uniform prior distribution to estimate the free energy difference.
To benchmark BayesBAR, we also computed the free energy difference using the BAR method and compared the results from both methods with the true value (Table \ref{tbl:twoh}).
The forces constants are set to be $k_1 = 25$ and $k_2 = 36$.
The number of samples, $n_1$ and $n_2$, are set equal and range from 10 to 5000.
For each sample size, we repeated the calculation for $K = 100$ times and computed the root mean squared error (RMSE), the bias, and the standard deviation (SD) of the estimates.
The RMSE is computed as $\sqrt{\sum_{k=1}^{K}(\Delta \hat{F}_i - \Delta F)^2/K}$, where $\Delta\hat{F}_k$ is the estimate
from the $k$th repeat and $\Delta F$ is the true value.
The bias is computed as $\Delta \bar{F} - F$, where $\Delta \bar{F} = \sum_{k=1}^{K}\Delta\hat{F}_i/K$, and 
the SD is computed as $\sqrt{\sum_{k=1}^K(\Delta \hat{F}_i - \Delta \bar{F})^2/(K-1)}$.

\begin{table}
  \caption{Free energy difference between the two harmonic oscillators ($k_1 = 25, k_2 = 36$).}
  \label{tbl:twoh}
  \small
  \begin{tabular}{ccccccccccc}
    \hline
    \multirow{2}{*}{$n_1 (=n_2)$} & \multicolumn{2}{c}{RMSE} & \multicolumn{2}{c}{Bias}  & \multicolumn{2}{c}{SD}& \multicolumn{4}{c}{Estimate of SD}   \\
    \cmidrule(l{0pt}r{1.5pt}){2-3} \cmidrule(l{1.5pt}r{1.5pt}){4-5} \cmidrule(l{1.5pt}r{1.5pt}){6-7} \cmidrule(l{1.5pt}r{0pt}){8-11}
    & MAP\textsuperscript{\emph{a}} & mean\textsuperscript{\emph{b}} & MAP & mean & MAP & mean & BayesBAR & asymptotic & Bennett's & bootstrap \\
    \hline
    10 &        2.45 &        2.40 &        0.85 &        0.90 &       2.29 &       2.23 &    4.08 &   39.24 &      1.10 &        1.53 \\
    13 &        2.53 &        2.47 &        0.87 &        0.92 &       2.38 &       2.29 &    3.55 &   19.69 &      1.11 &        1.47 \\
    18 &        1.92 &        1.84 &        0.16 &        0.22 &       1.91 &       1.83 &    3.09 &   11.58 &      1.09 &        1.31 \\
    28 &        1.72 &        1.65 &        0.44 &        0.48 &       1.66 &       1.58 &    2.58 &    7.06 &      1.07 &        1.14 \\
    48 &        1.27 &        1.19 &        0.16 &        0.19 &       1.26 &       1.17 &    1.90 &    2.92 &      1.07 &        1.11 \\
    99 &        1.34 &        1.23 &       -0.19 &       -0.11 &       1.33 &       1.22 &    1.38 &    1.64 &      0.98 &        0.96 \\
   304 &        0.83 &        0.79 &        0.00 &        0.03 &       0.83 &       0.79 &    0.80 &    0.81 &      0.74 &        0.70 \\
  5000 &        0.19 &        0.18 &       -0.00 &       -0.00 &       0.19 &       0.18 &    0.20 &    0.20 &      0.20 &        0.20 \\
    \hline
  \end{tabular}
  \textsuperscript{\emph{a}} Maximum a posteriori probability (MAP) estimate of BayesBAR (equivalent to the BAR estimate); \textsuperscript{\emph{b}} Posterior mean estimate of BayesBAR.
\end{table}

Because the uniform prior distribution is used, the MAP estimate of BayesBAR is identical to the BAR estimate.
Besides the MAP estimate, BayesBAR also provides the posterior mean estimate, which is computed using numerical integration.
Compared to the MAP estimate (the BAR estimate), the posterior mean estimate has a smaller RMSE.
Decomposing the RMSE into bias and SD, we found that the posterior mean estimate has a larger bias but a smaller SD than the MAP estimate.
The decrease in SD over compensates the increase in bias for the posterior mean estimate,  which leads to its smaller RMSE.
Statistical testing shows that the differences in RMSE and bias between the MAP estimate and the posterior mean estimate are statistically significant (p-value $< 0.05$), while the difference in SD is not (Figure S1 and S2).
Although the MAP estimate and the posterior mean estimate have different RMSEs, the difference is small and both estimates converge to the true value as the sample size increases.
This suggests that both estimates can be used interchangeably in practice.

Besides the MAP and the posterior mean estimate for $\Delta F$, BayesBAR offers an estimate of the uncertainty (whose true values are included in the two columns beneath the label SD in Table \ref{tbl:twoh}) using the posterior standard deviation 
(the BayesBAR column in Table \ref{tbl:twoh}).
For benchmarking, we also calculated the uncertainty estimate using asymptotic analysis, Bennett's method, and the bootstrap method.
Because each repeat produces an uncertainty estimate, we used the average from all $K$ repeats as the uncertainty estimate of each method, denoted as ``Estimate of SD'' in Table \ref{tbl:twoh}.

When the number of configurations is small, the asymptotic analysis significantly overestimates the uncertainty, while both Bennett's method and the bootstrap method tend to underestimate the uncertainty.
Practically, overestimating uncertainty is favored over underestimating, as the former prompts further configuration collection, whereas the latter might cause the user to stop sampling prematurely.
Nevertheless, excessive overestimation isn't ideal either, as it might result in gathering an unnecessarily large number of configurations.
Given these considerations, BayesBAR's uncertainty estimate overestimates the uncertainty modestly and thus is a better choice than the other methods.
As the sample size increases, the uncertainty estimates from all methods converge to the true value.

\begin{figure}
  \centering
  \includegraphics[width=0.8\textwidth]{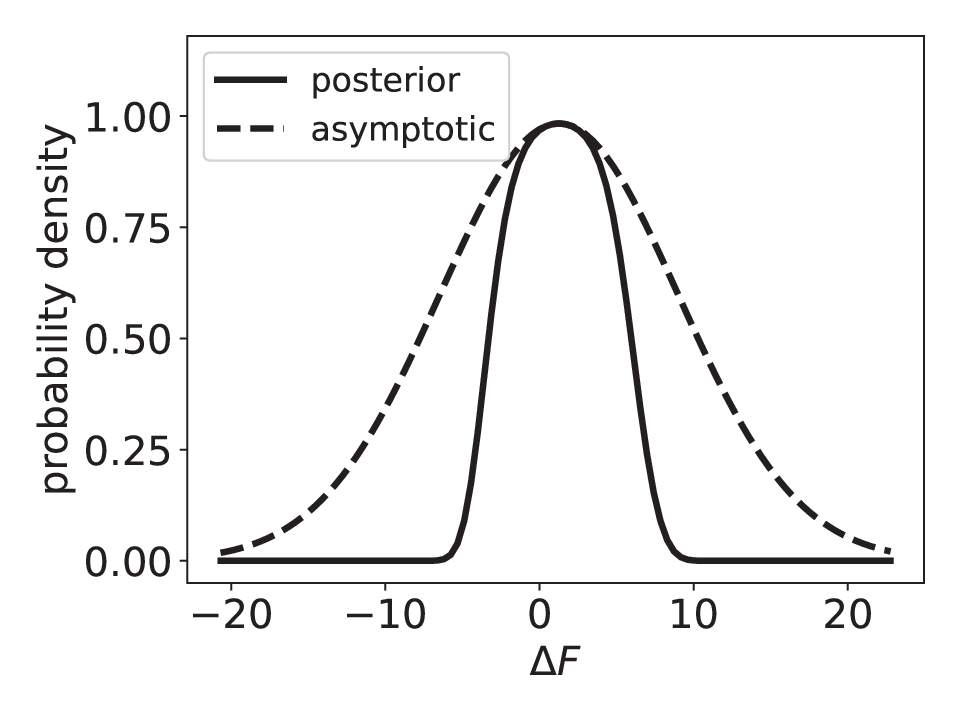}
  \caption{Probability densities of the posterior distribution (solid line) of $\Delta F$ and 
  the approximate Gaussian distribution (dashed line) used by the asymptotic analysis
  for the two harmonic oscillator system with $n_1 = n_2 = 18$.}
  \label{fig:twoh}
\end{figure}

The asymptotic analysis tends to overestimate uncertainty much more than BayesBAR. 
This is because the asymptotic analysis approximates the posterior distribution of $\Delta F$ with a Gaussian distribution centered around the MAP estimate. 
Such an approximation is generally accurate for a large number of configurations. 
However, with a smaller number of configurations, this approximation becomes imprecise, leading to considerable overestimation of uncertainty. 
Fig. \ref{fig:twoh} provides a visual comparison, contrasting the posterior distribution of $\Delta F$ as determined by BayesBAR with the Gaussian approximation from the asymptotic analysis for an experiment where $n_1 = n_2 = 18$.

\subsection*{Computing free energy differences among three harmonic oscillators.}
\begin{figure}
  \centering
  \includegraphics[width=0.8\textwidth]{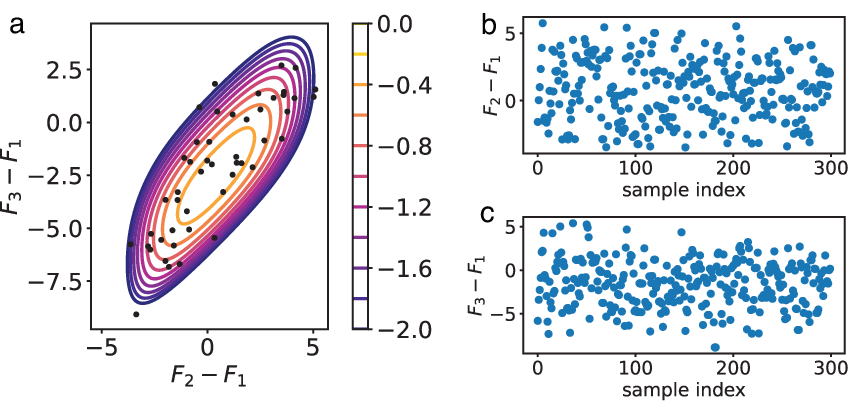}
  \caption{Probability density and samples of the posterior distribution of 
  $F_2 - F_1$ and $F_3 - F_1$ for the three harmonic oscillators with $n = 18$.
  (a) Contours are the logarithm of the posterior distribution density.
  Dots are a subset of samples drawn from the posterior distribution using the NUTS sampler.
  (b and c) The first 300 samples of $F_2 - F_1$ and $F_3 - F_1$ drawn from the posterior distribution using the NUTS sampler.}
  \label{fig:threeh}
\end{figure}

We next tested the performance of BayesMBAR on a multistate system.
The system consists of three harmonic oscillators with the following unitless potential energy functions:
$u_1(x) = \frac{1}{2} k_1 x^2$, $u_2(x) = \frac{1}{2} k_2 (x - 1)^2$, and $u_3(x) = \frac{1}{2} k_3 (x - 2)^2$, where $k_1 = 16$, $k_2 = 25$, and $k_3 = 36$.
The free energy differences among the three harmonic oscillators are analytically known.
Similar to the two harmonic oscillator system, we first draw $n$ samples form the Boltzmann distribution of each harmonic oscillator.
We use BayesMBAR with the uniform prior to estimate the free energy differences by computing both the MAP estimate and the posterior mean estimate.
The posterior mean estimate is computed by sampling from the posterior distribution using the NUTS sampler instead of numerical integration.
Figure \ref{fig:threeh} shows the posterior distribution of the free energy differences ($F_2 - F_1$ and $F_3 - F_1$) and a subset of samples drawn from the posterior distribution in one repeat of the calculation when $n = 18$.
As shown in Figure \ref{fig:threeh}(b) and \ref{fig:threeh}(c), samples from the NUTS samplers decorrelate quickly and  can efficiently traverse the posterior distribution.

For benchmarking purposes, we conducted the calculation 100 times ($K = 100$) for each sample size $n$, and derived metrics including the RMSE, bias, and SD of the estimate (Table \ref{tbl:threeh}). 
Given the use of a uniform prior, BayesMBAR's MAP estimate is the same as the MBAR estimate. 
When contrasted with the MBAR estimate, the posterior mean estimate has lower SD but higher bias. When factoring in both SD and bias, the posterior mean estimate has a smaller RMSE compared to the MBAR estimate. 
The differences in RMSE and bias are statistically significant when $n \leq 99$ and the difference in SD is not for any sample size (Figure S3-S6).
As in case of two harmonic oscillators, the difference in RMSE between the two estimators is minimal, and both estimates converges to the correct value as sample size grows.
In terms of uncertainty, BayesMBAR offers a superior estimate compared to established techniques like asymptotic analysis or the bootstrap method, especially with limited configuration sizes. 
Notably, BayesMBAR's uncertainty estimate avoids the underestimation seen with the bootstrap method. 
Simultaneously, compared to the asymptotic analysis, BayesMBAR's uncertainty estimate has a more modest overestimation.

\begin{table}
  \caption{Free energy differences among the three harmonic oscillators ($k_1 = 16, k_2 = 25, k_3 = 36$).}
  \label{tbl:threeh}
  \small
  \begin{tabular}{cccccccccc}
    \multicolumn{10}{c}{$F_2 - F_1$} \\
    \hline
    \multirow{2}{*}{$n$} & \multicolumn{2}{c}{RMSE} & \multicolumn{2}{c}{Bias}  & \multicolumn{2}{c}{SD}& \multicolumn{3}{c}{Estimate of SD}   \\
    \cmidrule(l{0pt}r{1.5pt}){2-3} \cmidrule(l{1.5pt}r{1.5pt}){4-5} \cmidrule(l{1.5pt}r{1.5pt}){6-7} \cmidrule(l{1.5pt}r{0pt}){8-10}
    & MAP & mean & MAP & mean & MAP & mean & BayesMBAR & asymptotic  & bootstrap \\
    \hline
    10 &        1.89 &        1.83 &        0.45 &        0.53 &       1.84 &       1.75 &    2.28 &    5.31 &        1.26 \\
    13 &        1.84 &        1.76 &        0.46 &        0.52 &       1.78 &       1.68 &    1.93 &    3.20 &        1.19 \\
    18 &        1.41 &        1.30 &       -0.09 &        0.01 &       1.41 &       1.30 &    1.62 &    2.17 &        1.05 \\
    28 &        1.18 &        1.12 &        0.13 &        0.19 &       1.17 &       1.10 &    1.31 &    1.55 &        0.91 \\
    48 &        0.79 &        0.75 &       -0.04 &        0.01 &       0.79 &       0.75 &    0.97 &    1.00 &        0.83 \\
    99 &        0.67 &        0.64 &       -0.06 &       -0.04 &       0.66 &       0.64 &    0.69 &    0.70 &        0.63 \\
   304 &        0.39 &        0.39 &       -0.01 &       -0.00 &       0.39 &       0.39 &    0.40 &    0.40 &        0.40 \\
  5000 &        0.09 &        0.09 &       -0.00 &        0.00 &       0.09 &       0.09 &    0.10 &    0.10 &        0.10 \\
    \hline
    \\
    \multicolumn{10}{c}{$F_3 - F_1$} \\
    \hline
    \multirow{2}{*}{$n$} & \multicolumn{2}{c}{RMSE} & \multicolumn{2}{c}{Bias}  & \multicolumn{2}{c}{SD}& \multicolumn{3}{c}{Estimate of SD}   \\
    \cmidrule(l{0pt}r{1.5pt}){2-3} \cmidrule(l{1.5pt}r{1.5pt}){4-5} \cmidrule(l{1.5pt}r{1.5pt}){6-7} \cmidrule(l{1.5pt}r{0pt}){8-10}
    & MAP & mean & MAP & mean & MAP & mean & BayesMBAR & asymptotic  & bootstrap \\
    \hline
    10 &        2.93 &        2.85 &        0.89 &        1.02 &       2.79 &       2.66 &    4.63 &   28.27 &        2.02 \\
    13 &        3.26 &        3.18 &        1.25 &        1.35 &       3.01 &       2.87 &    4.16 &   20.74 &        1.86 \\
    18 &        2.53 &        2.40 &        0.12 &        0.27 &       2.53 &       2.39 &    3.39 &   10.12 &        1.85 \\
    28 &        2.28 &        2.20 &        0.62 &        0.73 &       2.20 &       2.08 &    2.87 &    6.35 &        1.56 \\
    48 &        1.73 &        1.64 &        0.16 &        0.26 &       1.72 &       1.62 &    2.26 &    3.91 &        1.37 \\
    99 &        1.53 &        1.43 &        0.36 &        0.42 &       1.49 &       1.37 &    1.58 &    1.84 &        1.21 \\
   304 &        0.99 &        0.95 &       -0.00 &        0.03 &       0.99 &       0.95 &    0.89 &    0.91 &        0.81 \\
  5000 &        0.23 &        0.22 &        0.01 &        0.01 &       0.23 &       0.22 &    0.22 &    0.23 &        0.22 \\
    \hline
  \end{tabular}
\end{table}

\subsection*{Computing the hydration free energy of phenol.}
We further tested the performance of BayesMBAR on a realistic system that involves collective variables.
Specifically, we use BayesMBAR to compute the hydration free energy of phenol using an alchemical approach.
In this approach, we modify the non-bonded interactions between phenol and water using an alchemical variable $\lambda = (\lambda_\mathrm{elec}, \lambda_\mathrm{vdw})$, where $\lambda_\mathrm{elec}$ and $\lambda_\mathrm{vdw}$ are alchemical variables for the electrostatic and the van der Waals interactions, respectively.
The electrostatic interaction is linearly scaled by $1 - \lambda_\mathrm{elec}$ as 
\begin{align}
E_\mathrm{elec}(\lambda_\mathrm{elec}) = (1-\lambda_\mathrm{elec}) \frac{q_i q_j}{r_{ij}},
\end{align}
where $q_i$ and $q_j$ are the charges of atom $i$ and $j$, respectively; and $r_{ij}$ is the distance between them.
The van der Walls interaction is modified by $\lambda_\mathrm{vdw}$ using a soft-core Lennard-Jones potential\cite{beutlerAvoidingSingularitiesNumerical1994} as 
\begin{align}
E_\mathrm{vdw}(\lambda_\mathrm{vdw}) = (1-\lambda_\mathrm{vdw}) \cdot 4 \epsilon_{ij} \left[ \frac{1}{(\alpha \cdot  \lambda_\mathrm{vdw} + (r_{ij}/\sigma_{ij})^6)^2} - \frac{1}{\alpha \cdot  \lambda_\mathrm{vdw} + (r_{ij}/\sigma_{ij})^6} \right],
\end{align}
where $\epsilon_{ij}$ and $\sigma_{ij}$ are the Lennard-Jones parameters of atom $i$ and $j$; and $\alpha = 0.5$.
When $(\lambda_\mathrm{elec}, \lambda_\mathrm{vdw}) = (0, 0)$, the non-bonded interactions between phenol and water are turned on and phenol is in the water phase.
When $(\lambda_\mathrm{elec}, \lambda_\mathrm{vdw}) = (1, 1)$, the non-bonded interactions are turned off and phenol is in the vacuum phase.
The hydration free energy of phenol is equal to the free energy difference between the two states of $\lambda = (0, 0)$ and $\lambda = (1, 1)$.
To compute the free energy difference, we introduce 7 intermediate states through which $\lambda_\mathrm{elec}$ and $\lambda_\mathrm{vdw}$ are gradually changed from $(0, 0)$ to $(1, 1)$.
The values of $\lambda_\mathrm{elec}$ and $\lambda_\mathrm{vdw}$ for the intermediate and end states are included in Table \ref{tbl:lambda}.

\begin{table}[htbp]
  \centering
  \caption{The values of $\lambda_\mathrm{elec}$ and $\lambda_\mathrm{vdw}$ used in computing the hydration free energy of phenol.}
  \label{tbl:lambda}
  \begin{tabular}{l|rrrrrrrrr}
  \toprule
  $\lambda$ index & 1 & 2 & 3 & 4 & 5 & 6 & 7 & 8 & 9 \\
  \midrule
  $\lambda_\mathrm{elec}$ & 0 & 0.33 & 0.66 & 1 & 1 & 1 & 1 & 1 & 1 \\
  $\lambda_\mathrm{vdw}$ & 0 & 0 & 0 & 0 & 0.2 & 0.4 & 0.6 & 0.8 & 1 \\
  \bottomrule
  \end{tabular}
\end{table}

We use the general amber force field \cite{wangDevelopmentTestingGeneral2004} for phenol and the TIP3P model\cite{jorgensenComparisonSimplePotential1983} for water.
The particle mesh Ewald (PME) method\cite{dardenParticleMeshEwald1993}  is used to compute the electrostatic interactions.
Lennard-Jones interactions are truncated at 12 \AA\  with a switching function starting at 10 \AA.
We use OpenMM\cite{eastmanOpenMMRapidDevelopment2017} to run \textit{NPT} molecular dynamics simulations for all states at 300 K and 1 atm using the middle scheme\cite{zhangUnifiedEfficientThermostat2019} Langevin integrator with a friction coefficient of 1 ps$^{-1}$ and the Monte Carlo barostat\cite{aqvistMolecularDynamicsSimulations2004} with a frequency of 25 steps.
Each simulation is run for 20 ns with a time step of 2 fs.
Configurations are saved every 20 ps to ensure that they are uncorrelated \cite{choderaSimpleMethodAutomated2016} (Figure S7), as verified in the Supporting Information.
We use BayesMBAR to compute the free energy differences with $n$ configurations from each state.
We repeated the calculation for $K = 100$ times.
Because the ground truth hydration free energy is not known analytically, we use as the benchmark the MBAR estimate computed using all configurations sampled from all repeats, i.e., 100,000 configurations from each state.

\begin{figure}
  \centering
  \includegraphics[width=0.8\textwidth]{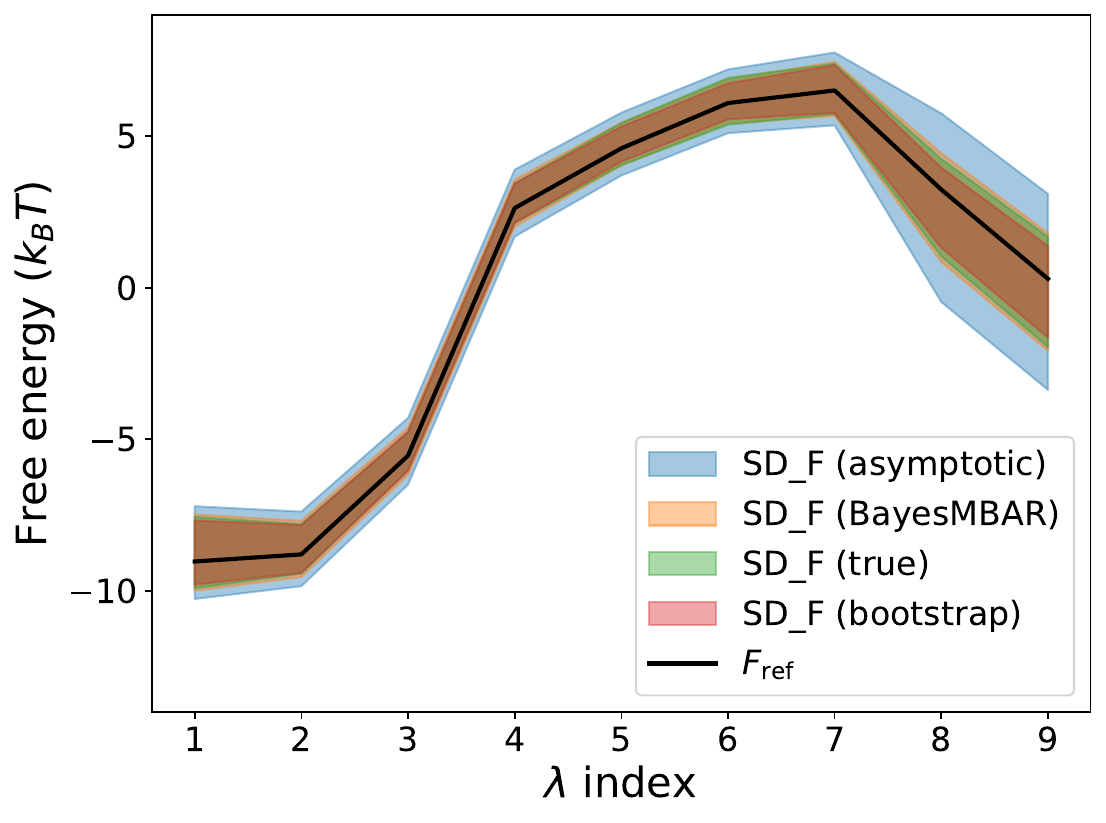}
  \caption{Free energy estimates of all states for computing the hydration free energy of phenol. 
  SD\textunderscore F (BayesMBAR), SD\textunderscore F (asymptotic), and SD\textunderscore F (bootstrap) are the average of the  uncertainty estimates using BayesMBAR, the asymptotic analysis, and the bootstrap method, respectively, when $n = 5$.
  SD\textunderscore F (true) is the true uncertainty when $n = 5$.
  $F_\mathrm{ref}$ is the MBAR estimate computed using all configurations sampled from all repeats.}
  \label{fig:phenol}
\end{figure}

\textbf{Uniform prior.} We first tested the performance of BayesMBAR with the uniform prior using different numbers of configurations.
Here the $n$ configurations from each state are not randomly sampled from saved configurations during the 20 ns of simulation.
Instead, we use the first $n$ configurations to mimic the situation in production calculations where configurations are saved sequentially.
The results are summarized in Table \ref{tbl:phenol}.
Compared to the MAP estimate (the MBAR estimate), the posterior mean estimate has a smaller SD but larger bias, as observed in the previous harmonic oscillator systems.
In terms of RMSE, the posterior mean estimate has a larger RMSE than the MAP estimate, which is different from that in the harmonic oscillator systems.
However, the differences in RMSE and bias between the MAP estimate and the posterior mean are not statistically significant except when $n = 5000$ (Figure S8) and the difference in SD is not statistically significant for any sample size (Figure S8).
This further suggests that both estimates can be used interchangeably in practice.
We also compared the uncertainty estimate among BayesMBAR, asymptotic analysis, and the bootstrap method.
The BayesMBAR estimate of the uncertainty is closer to the true value than the asymptotic analysis while not underestimating the uncertainty as the bootstrap method does when the number of configurations is small.
In addition to the free energy difference between the two end states, we also compared the uncertainty estimates for free energies of all states (Figure \ref{fig:phenol}).
When the number of configurations is small ($n = 5$), the uncertainty estimates from BayesMBAR are closer to the true uncertainty than the asymptotic analysis and does not underestimate the uncertainty as the bootstrap method does.

\begin{table}
  \caption{Hydration free energy (in the unit of $k_bT$) of phenol computed using BayesMBAR with the uniform prior.}
  \label{tbl:phenol}
  \small
  \begin{tabular}{cccccccccc}
    \hline
    \multirow{2}{*}{$n_1 (=n_2)$} & \multicolumn{2}{c}{RMSE} & \multicolumn{2}{c}{Bias}  & \multicolumn{2}{c}{SD}& \multicolumn{3}{c}{Estimate of SD}   \\
    \cmidrule(l{0pt}r{1.5pt}){2-3} \cmidrule(l{1.5pt}r{1.5pt}){4-5} \cmidrule(l{1.5pt}r{1.5pt}){6-7} \cmidrule(l{1.5pt}r{0pt}){8-10}
    & MAP & mean & MAP & mean & MAP & mean & BayesMBAR & asymptotic  & bootstrap \\
    \hline
    5 &        2.75 &        2.81 &       -0.73 &       -1.02 &       2.65 &       2.62 &        2.89 &       4.48 &     2.33 \\
    7 &        2.53 &        2.56 &       -0.60 &       -0.86 &       2.45 &       2.42 &        2.48 &       3.36 &     1.99 \\
   12 &        1.94 &        1.96 &       -0.23 &       -0.42 &       1.92 &       1.91 &        1.88 &       2.15 &     1.59 \\
   25 &        1.34 &        1.33 &       -0.03 &       -0.15 &       1.34 &       1.33 &        1.26 &       1.28 &     1.16 \\
   75 &        0.69 &        0.69 &        0.01 &       -0.04 &       0.69 &       0.69 &        0.73 &       0.73 &     0.71 \\
 1000 &        0.19 &        0.19 &       -0.00 &       -0.01 &       0.19 &       0.19 &        0.20 &       0.20 &     0.20 \\
    \hline
  \end{tabular}
\end{table}

\textbf{Normal prior.}
The free energy surface along the alchemical variable $\lambda$ is expected to be smooth, so we can use a normal prior distribution in BayesMBAR to encode this prior knowledge.
The squared exponential covariance function is defined as
\begin{align}
  k_{\textrm{SE}}(\lambda, \lambda^\prime) = \sigma^2 \exp\Big(-\frac{1}{2} \Big[ \frac{(\lambda_\mathrm{elec} - \lambda_\mathrm{elec}^\prime)^2}{l_\mathrm{elec}^2} + \frac{(\lambda_\mathrm{vdw} - \lambda_\mathrm{vdw}^\prime)^2}{l_\mathrm{vdw}^2} \Big] \Big),
\end{align}
where $\sigma^2$ is the variance and $l_\mathrm{elec}$ and $l_\mathrm{vdw}$ are the length scales for $\lambda_\mathrm{elec}$ and $\lambda_\mathrm{vdw}$, respectively.

The hyperparameters in the covariance functions and the mean parameter of the prior distribution are optimized by maximizing the Bayesian evidence.
After optimizing the hyperparameters, we use the MAP and the posterior mean estimators to estimate the free energy difference between the two end states and compare them to the MAP estimator with the uniform prior distribution (Table \ref{tbl:alchem_normal}), which is identical to the MBAR estimator.

\begin{table}
  \caption{Comparison of the performance of BayesMBAR with the uniform prior (MAP estimate) and the normal prior (MAP and posterior mean estimates) for computing the hydration free energy of phenol.}
  \label{tbl:alchem_normal}
  \small
  \begin{tabular}{cccccccccc}
    \hline
    \multirow{3}{*}{$n_1 (=n_2)$} & \multicolumn{3}{c}{RMSE} & \multicolumn{3}{c}{Bias}  & \multicolumn{3}{c}{SD} \\
    \cmidrule(l{0pt}r{1.5pt}){2-4} \cmidrule(l{1.5pt}r{1.5pt}){5-7} \cmidrule(l{1.5pt}r{1.5pt}){8-10} 
    & \multirow{2}{*}{uniform} & \multicolumn{2}{c}{normal} & \multirow{2}{*}{uniform} & \multicolumn{2}{c}{normal} & \multirow{2}{*}{uniform} & \multicolumn{2}{c}{normal}  \\
    \cmidrule(l{0pt}r{1.5pt}){3-4} \cmidrule(l{1.5pt}r{1.5pt}){6-7} \cmidrule(l{1.5pt}r{1.5pt}){9-10}
    & & MAP & mean & & MAP & mean & & MAP & mean \\
    \hline
    5 & 2.75 & 2.18 & 2.15 & -0.73 &  0.69 &  0.57 & 2.65 & 2.07 & 2.07 \\
    7 & 2.53 & 1.97 & 1.96 & -0.60 &  0.66 &  0.55 & 2.45 & 1.86 & 1.88 \\
   12 & 1.94 & 1.60 & 1.61 & -0.23 &  0.56 &  0.49 & 1.92 & 1.50 & 1.53 \\
   25 & 1.34 & 1.22 & 1.22 & -0.03 &  0.43 &  0.38 & 1.34 & 1.15 & 1.16 \\
   75 & 0.69 & 0.75 & 0.75 &  0.01 &  0.11 &  0.09 & 0.69 & 0.74 & 0.74 \\
 1000 & 0.19 & 0.20 & 0.20 & -0.00 & -0.02 & -0.02 & 0.19 & 0.19 & 0.19 \\     
    \hline
  \end{tabular}
\end{table}

By incorporating the prior knowledge of the smoothness of the free energy surface, the BayesMBAR estimator with a normal prior distribution has a smaller RMSE than the MBAR estimator, especially when the number of configurations is small.
As the number of configurations increases, the BayesMBAR estimator converges to the MBAR estimate.
When the number of configurations is small, the information about free energy from data is limited and the prior knowledge of the free energy surface excludes unlikely results and helps improve the estimate.
When the number of configurations is large, the inference is dominated by the data and the prior knowledge becomes less important, because the prior knowledge used here is a relatively weak prior.
This behavior is desirable because the prior knowledge should be used when data alone are not sufficient to make a good inference and at the same time not bias the inference when data are sufficient.

\section{Conclusion and Discussion}
In this study, we developed BayesMBAR, a Bayesian generalization of the MBAR method based on the reverse logistic regression formulation of MBAR.
BayesMBAR provides a posterior distribution of free energy, which is used to estimate free energies and compute the estimation uncertainty.
When uniform distributions are used as the prior, the MAP estimate of BayesMBAR recovers the MBAR estimate.
Besides the MAP estimate, BayesMBAR provides the posterior mean estimate of free energy.
Compared to the MAP estimate, the posterior mean estimate tends to have a larger bias but a smaller SD.
The reason for such observation could be that the posterior mean estimate takes into account the whole spread of the posterior distribution, which makes it more stable over repeated calculations and at the same time makes it more susceptible to extreme values.
The difference in accuracy between the MAP estimate and the posterior mean estimate is small and both estimates converge to the true value as the number of configurations increases.
Therefore both estimates can be used interchangeably in practice.
In BayesMBAR, the estimation uncertainty is computed using the posterior standard deviation.
All benchmark systems in this study show that such uncertainty estimate from BayesMBAR is better than that from the asymptotic analysis and the Bennett's method, especially when the number of configurations is small.

As a Bayesian method, BayesMBAR is able to incorporate prior knowledge about free energy into the estimation.
We demonstrated this feature by using a normal prior distribution to encode the prior knowledge of the smoothness of free energy surfaces.
All hyperparameters in the prior distribution are automatically optimized by maximizing the Bayesian evidence.
By using such prior knowledge, BayesMBAR provides more accurate estimate than the MBAR method when the number of configurations is small, and converges to the MBAR estimate when the number of configurations is large.

To facilitate the adoption of BayesMBAR, we provide an open-source Python package at \url{https://github.com/DingGroup/BayesMBAR}.
In cases where prior knowledge is not available, we recommend using BayesMBAR with the uniform distribution as the prior distribution. 
It takes the same input as MBAR and can thus be easily integrated into existing workflows with the benefit of providing better uncertainty estimates.
For computing free energy differences among points on a smooth free energy surface, we recommend using BayesMBAR with normal distributions as the prior.
Because the hyperparameters in the normal prior distribution are automatically optimized, the extra input to BayesMBAR from the user, compared to MBAR, is the values of the collective variables associated with each thermodynamic state.
In terms of covariance functions, although we used the squared exponential covariance function in this study, other covariance functions such as the Mat\'ern covariance function and the rational quadratic covariance function \cite{rasmussenGaussianProcessesMachine2005} could also be used.
The choice of covariance functions could also be informed by comparing the Bayesian evidence after optimizing their hyperparameters.

Because BayesMBAR needs to sample from the posterior distribution, it is computationally more expensive than MBAR that uses the asymptotic analysis to compute estimation uncertainty.
Table S1 shows the running time required by MBAR and BayesMBAR for computing the hydration free energy of phenol when $n = 1000$ configurations are used from each alchemical state.
Both MBAR and BayesMBAR were run on a graphic processing unit and the FastMBAR\cite{dingFastSolverLarge2019} implementation was used for MBAR calculations.
Considering that most of the computational cost for calculating free energies lies in sampling configurations from the equilibrium distribution and BayesMBAR provides better uncertainty estimates and more accurate free energy estimates when prior knowledge is available, we believe that the extra computational cost is worthwhile in practice.

BayesMBAR could also be extended to incorporate other types of prior knowledge about free energy, such as knowledge from other calculations or experimental data.
For example, when computing relative binding free energies of two ligands, A and B, with a protein using alchemical free energy methods, results from cheaper calculations such as docking or molecular mechanics/generalized Born surface area (MM/GBSA) calculations could be used as prior knowledge.
Specifically, the relative binding free energy is often calculated as $\Delta \Delta G^{\mathrm{binding}}_{\mathrm{A}\rightarrow \mathrm{B}} = \Delta G^{\mathrm{bound}}_{\mathrm{A}\rightarrow \mathrm{B}} - \Delta G^{\mathrm{unbound}}_{\mathrm{A}\rightarrow \mathrm{B}}$, where $\Delta G^{\mathrm{bound}}_{\mathrm{A}\rightarrow \mathrm{B}}$ and $\Delta G^{\mathrm{unbound}}_{\mathrm{A}\rightarrow \mathrm{B}}$ are free energy differences of changing ligand A to B alchemically in the bound and unbound states, respectively.
Computing $\Delta G^{\mathrm{unbound}}_{\mathrm{A}\rightarrow \mathrm{B}}$ is often much cheaper than computing $\Delta G^{\mathrm{bound}}_{\mathrm{A}\rightarrow \mathrm{B}}$ because the unbound state is in solvent and thus does not require simulations with the protein.
Therefore, $\Delta G^{\mathrm{unbound}}_{\mathrm{A}\rightarrow \mathrm{B}}$ can be efficiently computed to a high precision.
On the other hand, docking or MM/GBSA calculations could provide rough estimates on the absolute binding free energies of ligands A and B, whose difference provides an estimate $\mu$ of $\Delta \Delta G^{\mathrm{binding}}_{\mathrm{A}\rightarrow \mathrm{B}}$ and associated uncertainty $\sigma$, i.e., $\Delta \Delta G^{\mathrm{binding}}_{\mathrm{A}\rightarrow \mathrm{B}}$ has a normal distribution with mean $\mu$ and standard deviation $\sigma$.
Combining $\Delta G^{\mathrm{unbound}}_{\mathrm{A}\rightarrow \mathrm{B}}$ with the distribution of $\Delta \Delta G^{\mathrm{binding}}_{\mathrm{A}\rightarrow \mathrm{B}}$ from docking or MM/GBSA calculations, we could construct a normal distribution on $\Delta G^{\mathrm{bound}}_{\mathrm{A}\rightarrow \mathrm{B}} ( = \Delta G^{\mathrm{unbound}}_{\mathrm{A}\rightarrow \mathrm{B}} + \Delta \Delta G^{\mathrm{binding}}_{\mathrm{A}\rightarrow \mathrm{B}})$ and use it as the prior distribution when computing $\Delta G^{\mathrm{bound}}_{\mathrm{A}\rightarrow \mathrm{B}}$ with BayesMBAR.
Such estimated $\Delta G^{\mathrm{bound}}_{\mathrm{A}\rightarrow \mathrm{B}}$ using BayesMBAR could then be combined with $\Delta G^{\mathrm{unbound}}_{\mathrm{A}\rightarrow \mathrm{B}}$ to compute $\Delta \Delta G^{\mathrm{binding}}_{\mathrm{A}\rightarrow \mathrm{B}}$.
Rough estimates of $\Delta \Delta G^{\mathrm{binding}}_{\mathrm{A}\rightarrow \mathrm{B}}$ could also come from experimental data such as qualitative competitive binding assays that only provide whether ligand A binds better than ligand B.
In this case, the prior distribution of $\Delta \Delta G^{\mathrm{binding}}_{\mathrm{A}\rightarrow \mathrm{B}}$ could be a bounded uniform distribution with a lower or upper bound of 0.
Similarly, combining such prior knowledge on $\Delta \Delta G^{\mathrm{binding}}_{\mathrm{A}\rightarrow \mathrm{B}}$ with $\Delta G^{\mathrm{unbound}}_{\mathrm{A}\rightarrow \mathrm{B}}$, we could construct a bounded uniform distribution on $\Delta G^{\mathrm{bound}}_{\mathrm{A}\rightarrow \mathrm{B}}$ and use it as the prior distribution for computing $\Delta G^{\mathrm{bound}}_{\mathrm{A}\rightarrow \mathrm{B}}$ with BayesMBAR.
We believe that such extension of BayesMBAR could be useful in practice and will be explored in future studies.


\begin{acknowledgement}
The author thanks the Tufts University High Performance Compute Cluster that was utilized for the research reported in this paper.
\end{acknowledgement}

\begin{suppinfo}
Statistical tests comparing the MAP estimator and the posterior mean estimator of BayesMBAR using the uniform prior distribution for computing free energy differences,
saved configurations are uncorrelated in computing the hydration free energy of phenol,
the standard error of estimating the ensemble mean potential energy using saved configurations sampled from alchemical states of phenol,
running time of MBAR and BayesMBAR.
This information is available free of charge via the Internet at http://pubs.acs.org
\end{suppinfo}

\bibliography{BayesMBAR}

\end{document}